# Identifying Practical Challenges in the Implementation of Technical Measures for Data Privacy Compliance

*Completed Research Full Paper*

**Oleksandra Klymenko**
Technical University of Munich
alexandra.klymenko@tum.de

**Stephen Meisenbacher**
Technical University of Munich
stephen.meisenbacher@tum.de

**Florian Matthes**
Technical University of Munich
matthes@tum.de

## Abstract

Modern privacy regulations provide a strict mandate for data processing entities to implement appropriate technical measures to demonstrate compliance. In practice, determining what measures are indeed "appropriate" is not trivial, particularly in light of vague guidelines provided by privacy regulations. To exacerbate the issue, challenges arise not only in the implementation of the technical measures themselves, but also in a variety of factors involving the roles, processes, decisions, and culture surrounding the pursuit of privacy compliance. In this paper, we present 33 challenges faced in the implementation of technical measures for privacy compliance, derived from a qualitative analysis of 16 interviews with privacy professionals. In addition, we evaluate the interview findings in a survey study, which gives way to a discussion of the identified challenges and their implications.

## Introduction

The notion of data privacy has recently stepped into the spotlight as a result of two major movements: the prominence of big data (processing) and the regulatory reaction thereto. Fears of misuse, particularly regarding breaches of personal privacy, have spearheaded the rise of modern data protection regulations, such as The General Data Protection Regulation (GDPR), which serve as an attempt to curb the unregulated processing of personal information. The wide jurisdiction of the GDPR and its demonstrated enforceability (Wolff and Atallah 2020) have been noted, and its impact on organizations with a global reach is especially clear (Bennett 2018; Goddard 2017). The case for sound compliance is particularly strong; thus, safeguarding data is crucial to data processors.

The road to data privacy compliance begins with general recommendations from the GDPR and similar regulations to implement "appropriate technical and organisational measures." While this provides a starting point for privacy compliance, little detail is provided as to specific implementations and a general understanding of such measures does not fully exist (Klymenko et al. 2022). Nevertheless, attempts have been made to systematize technical measures (Huth and Matthes 2019; Piras et al. 2019). Such works, though, do not focus on technical measures; furthermore, a practical perspective is not investigated.

In this work, we aim to identify the practical challenges faced in the implementation of technical measures for data privacy compliance. In order to gain insight into the practical perspective, we interview 16 privacy professionals involved in the privacy compliance process. We then perform a thematic content analysis, yielding 33 challenges under four categories, which are validated in a survey study.

## Background and Related Work

Personal data must be handled in strict compliance with data protection laws and regulations such as The General Data Protection Regulation (GDPR), which is considered one of the strictest regulations to date. In their analysis of GDPR implementation and fines, Wolff and Atallah (2020) showed that the largest volume of fines, and largest average fines, were issued to organizations that implemented "insufficient technical and organizational measures" to qualify as being compliant.

Indeed, the text of GDPR often refers to the "appropriate technical and organizational measures" that must be implemented to be deemed compliant; however, it does not provide any specification as to what is considered *appropriate*. In the most general sense, technical measures can be understood as actions taken, with the help of certain technologies, to safeguard and preserve data privacy in some manner. In this context, of particular interest are Privacy-Enhancing Technologies (PETs), which are generally defined as "measures that protect privacy by eliminating or reducing personal data or by preventing unnecessary and/or undesired processing of personal data, all without losing the functionality of the information system" (Van Blarkom et al. 2003).

Works to identify challenges in protecting privacy post-GDPR adoption have been limited. Sirur et al. (2018) identify challenges in GDPR compliance through an interview study across multiple organizations, finding that data privacy compliance approaches can vary, especially in SMEs.

With respect to technical measures for privacy compliance, little research has been conducted to identify barriers to their implementation in practice. Piras et al. (2019) lead the way, stating that regulations like the GDPR provide little to no guidance on exact technical and organizational measures to be taken for compliance. Klymenko et al. (2022) support this notion, claiming that there is lack of understanding regarding the nature of technical measures. Similarly, Spiekermann (2012) calls the notion of Privacy by Design into question, in the way that such thinking might be difficult to integrate into existing systems.

The literature points to the existence of challenges regarding data privacy and compliance in modern technical systems, yet there is a lack of empirical evidence of the exact challenges relating to technical measures. Therefore, we strive to identify such challenges, systematize them, and evaluate their validity.

## Research Methodology

In this study, we aim to identify the practical challenges that currently exist in the implementation of technical measures for data privacy compliance. This is accomplished by interviewing practitioners from the technical and legal side of compliance. To guide this investigation, we define three research questions:

[RQ1] What are the practical challenges organizations encounter in their privacy compliance programs?
[RQ2] How can the identified challenges be organized and understood?
[RQ3] What is the prevalence of these challenges, as viewed by privacy professionals?

RQ1 investigates organizations' privacy compliance programs to identify what challenges exist, particularly with respect to technical measures. Next, these challenges are organized and analyzed, as per RQ2, to create a systematization of the challenges faced in the privacy compliance process. Finally, the challenges are validated empirically in a survey, where the prevalence of each challenge is tested.

### *Mixed Methods Study*

The research methodology follows a two-sided, mixed methods approach (Petter and Gallivan 2004; Venkatesh et al. 2013). Important to note is that the challenges to privacy compliance cannot be itemized a priori. Thus, the qualitative study follows a Grounded Theory approach (Glaser and Strauss 1967).

The qualitative study consists of semi-structured interviews of participants working with privacy in either the legal or technical sector. The interviews take inspiration from the framework of appreciative interviews (Cooperrider and Whitney 2006). A thematic content analysis (Braun and Clarke 2006) is conducted on the interview transcripts, which involves reviewing the transcripts, coding them, and categorizing codes into themes (constructs). This process of construct creation is iteratively performed concurrently to the interviews, following constant comparison in line with Grounded Theory Methodology (Wiesche et al. 2017). Finally, themes are validated by multiple coders to mitigate researcher bias.

The second phase of this study is carried out as a quantitative study in the form of surveys. The survey questions consist of statements relating to the identified challenges, with five "agreement" response options, taken from the Likert Scale: {strongly agree, agree, neither agree nor disagree, disagree, strongly disagree}. The deployment of this format is useful because it allows for simple aggregation at the end, which in turn enables one to evaluate the prevalence of the corresponding challenges.

**Instrumentation**

We developed an interview guide to be used in the semi-structured interviews. The goal of these questions was to gain insight into the roles and responsibilities of the interviewees, and therein, to identify and discuss challenges faced. Following the interviews and analysis, a survey was created using Google Forms. In the survey, each identified challenge was mapped to one or more survey questions.

**Interview and Survey Demographics**

16 semi-structured interviews were conducted, 9 from the legal domain and 7 from the technical. The interviewees were based in three different continents and worked at organizations ranging from micro- to large-sized. The relevant information of interviewees is summarized in Table 1. IDs suffixed with a "T" indicate a technical interviewee, "L" a legal, and "LT" one that traverses both fields. "Exp." denotes years of experience in the privacy field, and "Dur." indicates the duration of the interview in minutes. The survey study elicited 25 responses, 13 who reported working in a legal role and 12 in a technical. 28% of respondents have 10+ years of experience, while 28% had 0-3 years of experience. 76% of respondents came from large-sized organizations (over 200 employees).

| ID | Position | Organization | Exp. | Dur. |
|---|---|---|---|---|
| I1-T | Privacy Engineer | Large US media conglomerate | 10+ | 54 |
| I2-T | Privacy / Security Architect | Large German multinational software corporation | 6 | 52 |
| I3-L | Privacy / Security Lawyer | US law firm | 20+ | 32 |
| I4-T | Privacy Engineer | Large US multinational tech company | 5+ | 70 |
| I5-LT | DPO, Managing Director | German data protection software company | 4 | 50 |
| I6-T | Software Architect | Large German tech conglomerate | 3 | 50 |
| I7-L | Lawyer / External DPO | Small German data privacy company | 20+ | 55 |
| I8-L | Data Protection Counsel | International fintech corporation | 6 | 60 |
| I9-T | Privacy Engineer | Large US Tech Corporation | 8 | 65 |
| I10-LT | Legal Counsel | Global Web Consortium | 25 | 60 |
| I11-L | Legal Counsel | German-based privacy consulting firm | 3 | 50 |
| I12-L | DPO | German-based consulting firm | 20 | 55 |
| I13-T | Security / Privacy Architect | Large German tech conglomerate | 3 | 60 |
| I14-L | Compliance Officer | British-based news corporation | 3 | 50 |
| I15-T | Privacy Engineer | Large multinational Chinese tech corporation | 15 | 60 |
| I16-L | Legal Associate | Indian-based law firm | 3 | 55 |

**Table 1. Interview Study Participants**

# Qualitative Study Results

In this section, we introduce each identified challenge, broken down into four categories. Chief among the insights brought to light is that the factors influencing the implementation of technical measures include, yet extend beyond, the technologies themselves, revealing a unique dynamism and interplay.

### *The Technical-Legal Interaction*

With the implications of rising attention paid to data privacy, the technical-legal interaction has become more necessary (and commonplace) than ever. Because the two fields in general possess inherent differences, one can reasonably expect that a harmonization does not come immediately.

*C1.1: Rare Interaction* – Challenges begin with the cases in which this interaction happens rarely, or not at all. In a process where interdisciplinary communication can most likely come as a benefit to all parties involved, the lack of this dynamic can create challenges in mutual understanding.

*C1.2: Sub-optimal Interaction* – Technical-legal interaction may indeed occur, but in a somewhat sub-optimal manner. An added complexity comes with the question of whether *more* interaction is desired.

*C1.3: Technical-Legal Deadlock* – This challenge highlights a unilaterally perceived *deadlock* between technical and legal mindsets in the process of privacy compliance, as engineers and lawyers must essentially come to the same table of privacy compliance, and "speak the same language."

*C1.4: Lack of Domain-specific Knowledge* – Domain-specific knowledge is required for data privacy compliance, keeping in mind the very interdisciplinary nature of this field. Lacking knowledge from experts of differing backgrounds can lead to complications in the dynamic interactions that take place.

*C1.5: Lack of Technical Input* – This challenge refers to a perceived lack of technical input, i.e., input from technical experts, in the discussions surrounding compliance and its requirements. As such, it extends beyond the technologies themselves to the ecosystem of privacy compliance.

*C1.6: The Technical-Legal Gap, An Opposing View* – Implicit in this challenge is the acknowledgement that the gap between technical and legal experts is indeed "fundamental" or even necessary, since two different fields of study (and practice) are brought together within the process of privacy compliance.

*C1.7: Lack of Interdisciplinary Teams* – The existence of interdisciplinary units in the privacy compliance process, although mentioned in some interviews, is by no means standard. The challenges faced when people of different disciplines work together suggest the need to facilitate interdisciplinary interaction.

*C1.8: Interpretation of Regulations* – C1.8 is grounded in the idea that the makeup of today's most predominant data privacy regulations may, however inadvertently, create challenges for the implementation of technical measures, particularly in the question of interpretation of requirements.

## *Technologies*

In light of the technical measures in question, the second challenge category looks at the technology itself, and moreover, some existing barriers to the adoption and implementation of them. Specifically, an emphasis is placed on Privacy-Enhancing Technologies (PETs).

*C2.1: Lack of Proper Characterization* – This is aimed directly at the nature of PETs, in what is called a lack of "proper characterization." Such a notion sums up that while the motivation behind PETs is sound, its usability, understandability, and relevance to regulations are not quite clear or well-defined.

*C2.2: The Factor of Resources* – When introducing PETs to existing systems, the factor of resources cannot be ignored. This is particularly the case with PETs, which require resources in the form of infrastructure, but also through the education or hiring of experts to oversee their adoption.

*C2.3: Difficult to Communicate about PETs* – C2.3 focuses particularly on the complicated nature of PETs and their ability to be communicated effectively. Implicit is also the ease with which such technologies can be communicated between technical and legal experts, harking back to previously discussed challenges.

*C2.4: No Clear Mapping of PETs* – This challenge makes concrete the notion that there is no clear mapping of PETs to regulatory requirements. In addition, as alluded to in C2.1, preliminary work on the characterization of PETs does not generally include strong ties of their applicability to legal requirements.

*C2.5: Lack of Education on PETs* – A significant barrier to the concept of PETs is the general lack of knowledge, or lack of education on them. Extending beyond *knowledge*, C2.5 points to the underlying *implications* of such technologies. Thus, not only does *education* become important, but also *awareness*.

*C2.6: Lack of Technical Literacy* – In contrast to C1.4, this challenge focuses on how a lack of technical literacy affects the implementation of technical measures. Thus, the technical aspect is highlighted, particularly in relevance to regulations.

*C2.7: Lack of a Technical Framework* – C2.7 refers to the lack of an accepted and widely-used technical framework for the implementation of technical measures for privacy compliance. By "framework" is meant a structured technical *specification* to implement privacy-compliant systems.

*C2.8: No Incentive for PETs* – C2.8 addresses the notion that organizations striving to implement PETs are not "rewarded," i.e., there is no tangible incentive to exceed the bare minimum and invest in PETs.

*C2.9: Lack of Awareness of Existing Technologies* – Although the technology for data privacy protection may "be there," the crux of the matter becomes the people and organizations behind them. In essence, how promising "technical measures" are becomes irrelevant if the *awareness* of them is lacking.

*C2.10: The Privacy-Innovation Conundrum* – C2.10 is centered around the effect of privacy compliance on the innovation of technology. The challenge is that the process of privacy compliance does not necessarily promote innovation, which is referred to as the Privacy-Innovation Conundrum (Zarsky 2015).

## *Organizational Factors*

The third challenge category investigates the way in which the structure and culture of an organization can affect the implementation of technical measures for privacy compliance. Ultimately, to what extent sound technical measures are introduced is largely dependent on the people behind these decisions.

*C3.1: The Role of Management* – Although people serving in management positions may not be performing the actual implementation of technical measures, the decisions made at this level directly influence what procedures are put into place. Thus, the challenge exists in this dependence on management, placing a considerable emphasis on the roles at this level.

*C3.2: Dependence on Organizational Culture* – C3.2 posits that the implementation of technical measures is not performed in a vacuum; rather, the *organizational culture* surrounding them must be formed and made conducive to the willful and informed undertaking of such processes.

*C3.3: Lack of Uniform Structures and Roles* – C3.3 is *comparative* with respect to organizations and their privacy compliance structures. The argument here is that a lack of uniformly defined structures and roles as they pertain to privacy compliance can create challenges for the field moving forward.

*C3.4: Lacking Focus on Privacy Engineering* – C3.4 states that a larger emphasis needs to be placed on the practice of Privacy Engineering in privacy compliance, particularly with *technical* measures.

*C3.5: Varying Availability of Legal and Supervisory Support* – Within C3.5 becomes apparent the supportive role of legal entities and regulatory authorities. A particular gap comes to light concerning the overall availability of such authorities, something which seemingly has not been adequately researched.

*C3.6: The "It Depends" Problem* – This challenge resides purely in the industry, and how the definition of technical measures might differ from organization to organization. In short, it strongly depends on the data being processed, but also on a variety of organizational factors which may vary between entities.

*C3.7: Technical Measures as a Financial Concern* – Technical measures can become a financial concern in many steps along the way towards compliance, including the risk of fines, the resources to implement sound technologies, and the business value of being compliant.

*C3.8: Dependency on Risk Assessments* – C3.8 looks at the way in which the implementation of technical measures depends on a risk assessment that is conducted. The challenge here comes with the subjective nature of a risk assessment, as well as when such an assessment is not performed at all.

## *General*

The final challenge category is one that aims to tackle some of the broader challenges of privacy compliance, rather than pertaining specifically to technology or organizations.

*C4.1: Many Regulations and Settings* – C4.1 roots itself in the countless "settings" in which regulations must be considered and enforced, a discussion started by C3.6. These ideas illustrate the complexity of privacy compliance, and its implications for technical measures is an interesting point of study.

*C4.2: Bureaucracy* – C4.2 expresses that the process of privacy compliance involves a significant amount of bureaucracy, which can serve as a detriment to data privacy protection. Although difficult to quantify, the issue of bureaucracy can imaginably become a hindrance to compliance processes.

*C4.3: Industrialization of Privacy Compliance* – The term *industrialization* in this context refers to the shift to an industry of organizations that have been building software around privacy compliance. Although presented as a challenge, others could perceive this as a positive and beneficial development.

*C4.4: Inequality in Privacy Compliance* – The premise of C4.4 is rooted in the term of *bare minimum technical measures*, which points to the notion that demonstrating privacy compliance via technical measures should not be conflated with the protection of privacy.

*Practical Challenges of Technical Measures for Privacy Compliance**C4.5: What is Privacy?* – Privacy can be interpreted from a theoretical, personal, or even philosophical standpoint, and an added complexity comes with *data privacy*. Such complexity can lead to challenges in privacy *compliance*, particularly with the requirement of technical measures.

*C4.6: Conflating Privacy, Data Protection, and Security* – Privacy is often conflated with security, when it should be treated as distinct. Furthermore, the distinction between data privacy and data protection may not always be apparent. Such debates can have direct implications on the structure of privacy compliance programs, as the choice of technical measures rests upon the operational definition of privacy.

*C4.7: Losing the Meaning of Privacy* – The requirement of compliant data handling via the establishment of proper technical (and organizational) measures has been the de facto way in which data privacy can be safeguarded. As a result, a complex network of regulations, processes, and technologies has been constructed to form what has become the backbone of modern privacy compliance. In this dynamic, the challenge arises when the focus shifts away from privacy itself.

## Quantitative Study Results

The survey was created by mapping each identified challenge from the qualitative analysis to one or two survey statements. Square brackets in a statement indicate wording changes based on interviewee background. The survey responses are enumerated in Table 2, coded from Strongly Agree (SA) to Strongly Disagree (SD). The number in bold denotes the highest selected response per statement. In some cases, totals do not add up to the number of participants (25) due to the presence of a "N/A" option.

| Code | Survey Statement | SA | A | N | D | SD |
|---|---|---|---|---|---|---|
| C1.1 | I rarely interact with more [technically-\|legally-]oriented people regarding privacy compliance. | 2 | 4 | 3 | **7** | 7 |
| C1.2 | Interacting with [technical\|legal] experts about privacy matters can be slow or frustrating. | 2 | 3 | 7 | **9** | 0 |
| C1.3 | When interacting with [technical\|legal] experts about privacy matters, I feel like there is a disconnect that creates challenges. | 1 | **4** | 1 | 2 | 1 |
|  | There is a need for more/better interaction between the technical and legal sides of compliance. | 6 | **7** | 6 | 3 | 1 |
| C1.4 | I believe that in general, there is a lack of [technical\|legal] knowledge on the [legal\|technical] side. | **12** | 4 | 4 | 3 | 0 |
| C1.5 | Data privacy and privacy compliance are becoming more and more technically-centered. | 4 | **12** | 5 | 1 | 1 |
|  | There is a need for more technically-minded people in the conversation about privacy compliance. | 5 | **12** | 5 | 1 | 0 |
| C1.6 | Any perceived gap in knowledge or understanding between the technical and legal sides of privacy compliance is a good thing. | 1 | 5 | 5 | **10** | 2 |
| C1.7 | There is a lack of interdisciplinary/cross-functional teams in privacy compliance programs, i.e., a better balance is needed. | 6 | **11** | 3 | 2 | 1 |
| C1.8 | I feel like the makeup of current privacy regulations leaves much of the interpretation up to me. | 5 | **9** | 5 | 4 | 0 |
|  | I believe my answer to the question above is the optimal state of things. | 1 | **8** | 5 | 5 | 4 |
| C2.1 | I feel like there is a lack of proper characterization of PETs (function, benefits, disadvantages, etc.). | 2 | **12** | 5 | 3 | 0 |
|  | PETs are actually quite difficult to implement in practice. | 2 | 7 | **10** | 1 | 0 |
| C2.2 | The ability to implement PETs is very dependent on a company's resources. | 6 | **10** | 4 | 1 | 0 |
| C2.3 | In general, the understanding of PETs requires a sound technical baseline. | 4 | **11** | 4 | 2 | 0 |
|  | Communicating about these PETs with people with purely legal background is quite difficult. | 3 | 6 | **8** | 3 | 1 |
| C2.4 | There is no clear sense of how PETs relate to privacy regulations. | 0 | 6 | 6 | **7** | 1 |
| C2.5 | I believe there is a general need for better education on PETs. | 6 | **12** | 1 | 2 | 0 |
|  | I personally am interested in learning more about PETs. | 7 | **10** | 4 | 2 | 0 |
| C2.6 | There is a general lack of technical literacy when it comes to privacy compliance. | 3 | **10** | 4 | 5 | 1 |
| C2.7 | There does not exist a solid technical framework for privacy compliance. | 0 | **10** | 6 | 6 | 1 |
| C2.8 | I believe there is little incentive to put resources into implementing the newest, state-of-the-art Privacy-Enhancing Technologies. | 3 | 5 | **7** | 6 | 2 |
|  | It is more convenient and/or economical not to put time and resources into implementing sound technical measures. | 0 | **10** | 4 | 4 | 5 |
| C2.9 | I believe that there are many technologies for data protection, yet the awareness and knowledge surrounding them is lacking. | 2 | **15** | 2 | 4 | 0 |
| C2.10 | Privacy compliance doesn't motivate the innovation of technology in the privacy field. | 1 | 4 | 2 | **13** | 3 |
| C3.1 | When it comes to interpreting privacy regulations, this interpretation comes from the management level within my organization. | 2 | 1 | 7 | **10** | 3 |
| C3.2 | There are many ways to approach privacy compliance, specifically regarding technical measures. | 7 | **13** | 2 | 1 | 0 |
|  | There is an inequality in the current industry as to what suffices as being compliant. | 5 | **11** | 6 | 1 | 0 |
|  | Politics in an organization may affect the degree to which sound technical measures are pursued. | 9 | **10** | 3 | 1 | 0 |
| C3.4 | More focus should be placed on privacy engineering within organizations. | 8 | **11** | 3 | 1 | 0 |
| C3.5 | I would say legal support is readily available to me regarding any privacy-related matters. | 5 | **14** | 4 | 0 | 0 |

*Twenty-ninth Americas Conference on Information Systems, Panama, 2023* 6

| Code | Statement | | | | | |
|---|---|---|---|---|---|---|
| C3.6 | The implementation of technical measures for privacy compliance can be challenging depending on what systems are involved. | 8 | **13** | 1 | 1 | 0 |
| | Privacy compliance can be a bit of a gray area. | 3 | **16** | 3 | 1 | 0 |
| C3.7 | When it comes to technical measures for privacy compliance, it largely becomes a financial matter. | 1 | **10** | 5 | 6 | 1 |
| | Demonstrating sound privacy compliance can actually boost a company's value. | **12** | 8 | 2 | 0 | 1 |
| C3.8 | At the core of implementing technical measures for privacy compliance is a risk assessment, which is dependent on organizational culture. | 5 | **13** | 4 | 1 | 0 |
| C4.1 | I believe that the amount of privacy regulations nowadays makes implementing technical measures for privacy compliance convoluted. | 0 | 7 | **10** | 6 | 0 |
| | There is a clear need for better harmonization of these regulations (and requirements). | 3 | **11** | 7 | 2 | 0 |
| C4.2 | Privacy compliance involves too much bureaucracy. | 5 | 6 | 5 | **7** | 0 |
| C4.3 | All in all, I think recent years have seen the industrialization of privacy (compliance). | 6 | **11** | 6 | 0 | 0 |
| | This industrialization is in my view a positive advancement. | 5 | **11** | 4 | 2 | 1 |
| C4.4 | It is often the case with privacy compliance that although compliance can be argued for, true protection of privacy may not be achieved. | 5 | **10** | 5 | 3 | 0 |
| | There is such a thing as over-compliance. | 4 | **7** | 6 | 4 | 2 |
| C4.5 | In general, I think the concept of privacy itself is quite vague or not well-defined. | 2 | 2 | 7 | **9** | 3 |
| C4.6 | The meaning of privacy is distinct from, and often conflated with, the concept of data protection. | 4 | **8** | 4 | 6 | 1 |
| | Being 100% compliant does not imply 100% data protection. | **9** | 7 | 4 | 3 | 0 |
| C4.7 | The true meaning of privacy has been lost in modern privacy compliance. | 4 | 5 | **7** | 6 | 1 |

**Table 2. Complete Survey Results**

# Discussion

The ensuing discussion outlines the identified challenges in each category, grounds them in existing literature, and analyzes the survey results in light of the interview findings.

## *Exploring the Technical-Legal Divide*

The acknowledgment that there exist inherent differences between the technological and legal fields is not new. Gifford (2007) describes the relationship between law and technology as "both simple and exceedingly complex." In this dynamic, the value of domain-specific knowledge is brought to light. The study of this sort of domain-specific knowledge has certainly been investigated, particularly on the legal side. As it turns out, very few law schools offer courses on technical subjects (Ryan 2021), and this could be contributed to some of the identified challenges in this work.

One can also look to the nature of regulations as the source of challenges. Looking at these challenges requires first looking at the nature of *technical measures*. This phrase was specifically motivated by its use in the GDPR, and it is used generally to define the required measures in order to safeguard data and protect privacy, and ultimately to demonstrate compliance. The vagueness of GDPR as it pertains to technical measures is acknowledged in research, such as in (Mohan et al. 2019) or (Lindqvist 2018).

The survey results indicate general agreement with many of the challenges in this category. Among these, the importance of technical input in privacy compliance is emphasized, as well as the necessity of interpretation in the implementation of technical measures. The agreement also points to a need for better interaction between the technical and legal sides, possibly in the form of interdisciplinary teams. This is supported by works such as (Altman et al. 2021), and in general the theme of "bridging the gap" has received attention (Nissim et al. 2018).

## *Technical Measures as Technology*

As early as 1997, the utilization of Privacy-Enhancing Technologies was called for, specifically in light of increasing privacy concerns at the dawn of the Internet (Goldberg et al. 1997). The usability and understandability have been called into question (Renieris et al. 2021), and this has implications for privacy compliance. Attempts to characterize PETs have been undertaken (Huth and Matthes 2019; Heurix et al. 2015), but the specific relation to privacy regulation is not expounded upon.

The factor of resources presents a particular challenge, and this notion is confirmed by a recent report on PETs published by the Royal Society (Noble et al. 2019), which lists among the major limits of these technologies the requirement for vast computational resources and specialist skills. As noted by the challenges, another barrier comes with the knowledge and skills needed to harness these technologies, and a clear lack has been noted (Fischer-Hübner and Lindskog 2001), not only in university curriculum, but also in coverage of privacy and PETs in textbooks and learning material. An important point is made by Phillips: "Technical systems of this sort [PETs] are only as socially relevant as they are well incorporated into everyday practice" (Phillips 2004).

The pursuit of a technical framework for privacy compliance has been undertaken (Piras et al. 2019; Alshammari and Simpson 2018), yet the interviews in this study indicate that none have been uniformly accepted or adopted. This finding is also supported by the agreement observed in the survey.

Looking to the survey results, agreement with many of the challenges was indicated, particularly those alluding to the complexity of PETs, as well as the lack of technical foundation in compliance processes. The most resounding agreement came with the statement that while technologies do exist for compliance, awareness surrounding them is lacking. On the other hand, the sentiment that privacy compliance hinders innovation in privacy technologies was largely disagreed with.

### *Organizational Factors*

The influence of organizational culture on compliance practices has also been studied in the literature. Specifically pertaining to security compliance, Guenther claims that "the problem is not so much with security technology as it is with the lack of security awareness" (Guenther 2004). This can logically be extended to privacy. Similarly, research attention (Bellman et al. 2004; Benjamin 2017) has investigated the (socio-)cultural differences when it comes to views on information and data privacy.

An interesting finding comes with the influence of the field of Privacy Engineering in the implementation of technical measures for data privacy compliance. Gürses and del Alamo (2016) address a number of potential benefits from the emerging field of Privacy Engineering, among them the way in which it "responds to this gap between research and practice."

Also of note is the perceived challenge relating to the risk assessment that is intertwined with the implementation of technical measures, which is validated by the majority of agreement in the survey. Interestingly, the risk assessment carried out for privacy compliance can be viewed as the "common denominator" (Rose 2019), allowing organizations to evaluate their potential privacy risks and the strength of technical safeguards in place. The discrepancy in these perspectives can serve as a starting point for further investigations.

Many of the challenges presented in the survey were met with agreement, particularly with the challenge of privacy compliance often being a gray area, and accordingly the perceived inequality in compliance. A finding which challenges the idea that privacy compliance can be a financial burden comes with the strong agreement that it can actually boost a company's value. This, however, must be taken into consideration with the fact that compliance processes can still require a significant investment upfront.

### *Above and Beyond Technical Measures*

Challenges in the fourth category highlight that factors affecting privacy compliance programs extend above and beyond technical measures to issues and concerns with a wider scope. A particularly positive survey response came with the statement that being compliant does not necessarily equate to optimal data protection. As the challenges in this category lie on the edge of the scope of our study, we pose their further analysis and validation as a point for future investigation.

## Conclusion

In this work, we identify 33 challenges in the implementation of technical measures for data privacy compliance. These challenges can be divided into four categories, which clearly show the multi-faceted and complex nature of implementing technical measures. This categorization presents a novel approach of systematizing existing challenges in the process of implementing technical measures for privacy compliance. In order to begin the process of validating these challenges for relevance to industry, we conducted a survey study to quantify the perceived prevalence of the identified challenges. The results show a large majority of agreement, preliminarily verifying that many of the challenges do indeed occur in practice. Ultimately, our results can be seen as a roadmap for defining future areas of improvement.

A threat to internal validity arises from our methodology, which rests upon the findings of the thematic content analysis. Similarly, the main threat to external validity relating to our conducted study comes with the sample chosen for both the interviews and the survey. To mitigate these threats, we followed a defined methodology for the selection of study participants, which sought to facilitate the generalizability of our study findings. Nevertheless, more survey responses are needed to validate our results more completely.

The implications of our findings for practitioners are clear. As these challenges are derived from privacy professionals in their involvement with compliance programs, they represent barriers to more efficient compliance processes. As such, these challenges are practical problems requiring practical solutions.

In a similar vein, the implication for researchers first involves a call for a clearer understanding of the data privacy compliance ecosystem as a whole. Without this, one cannot begin to pursue further investigations into the challenges and their potential solutions. In addition, a sound research foundation into privacy compliance and its challenges can foster excellent industry-academic ties, with the opportunity for interdisciplinary collaboration to develop innovative solutions.

Therefore, we propose three streams of continuing work: (1) achieving a better understanding of the specific processes in place for privacy compliance, (2) targeted investigations into each challenge and an exploration of potential solutions, and (3) fostering interdisciplinary interactions to support these compliance processes. These present exciting opportunities to explore a field of increasing importance.

## Acknowledgements

This work has been supported by the German Federal Ministry of Education and Research (BMBF) Software Campus grant LACE 01IS17049.